# The plasmoelectric effect: optically induced electrochemical potentials in resonant metallic structures

Matthew T. Sheldon and Harry A. Atwater

*Thomas J. Watson Laboratories of Applied Physics, California Institute of Technology, MC 128-95, Pasadena, California 91125, USA*

**We describe a strategy for conversion of optical power into DC electrical power using resonant absorption in plasmonic nanostructures. A thermodynamic analysis of the underlying mechanism motivates our description of the phenomenon, which we term the plasmoelectric effect. Power conversion results from the dependence of optically generated heat on shifts of the plasmon resonance frequency that occur with changes of electron density. We model an all-metal device constructed from 10 nm radius silver spheres and predict a characteristic conversion efficiency of 14.3% under 1 kW m$^{-2}$ intensity, single-frequency radiation. We discuss strategies for enhanced efficiency, broadband power conversion, and further applications of this new class of optoelectronic device**.

Plasmonic nanostructures display remarkable optical properties arising from the coupling of their resonant free electron oscillations to incident light. Plasmonics has prompted significant scientific activity over the past decade, because plasmon excitation and propagation can be tailored by nanoscale control of size, shape and architecture [1]. Metal nanostructures exhibiting subwavelength optical confinement [2], enabling nanoscale photonic circuits, concentration schemes for photovoltaics [3, 4], field enhancement for Raman spectroscopy, and novel biological labeling techniques [5, 6], have been reported. Such coupling also results in excitation of extremely high energy densities during irradiation. Strong polarization occurs, with laser excitation inducing local charge density fluctuations of $\pm 10,000$ % or greater during each optical cycle [7]. Highly localized heating also occurs. Our calculations estimate a steady-state temperature increase of 100 K for resonant optical excitation of a 10 nm radius silver sphere at a power density of 1 kW m$^{-2}$. However, there is to date less understanding of how to effectively harness this energy. Notably, optoelectronic devices that emit optically excited hot electrons from plasmonic resonators across a rectifying semiconductor or insulator barrier into a circuit, have received attention as an energy generation mechanism, albeit with low efficiency [8-10]. Fast electronic relaxation via electron-phonon coupling in metals poses challenges to advancing the efficiency of hot-carrier collection in these devices [11].

In this Letter we describe the plasmoelectric effect and indicate a route to generation of DC electrical power from resonant optical absorption in plasmonic nanostructures. Rather than separation of optically excited electron-hole pairs, plasmoelectric power conversion results from the dependence of the plasmon resonance frequency, $\omega_p$, on electron density, $n$. As illustrated in Fig. 1, two metallic nanostructures, each with a different $\omega_p$, are electrically connected. The pair of resonators is irradiated at a frequency intermediate between the plasmon resonance frequencies $\omega_p$ for each individual neutral particle. Charge transport is thermodynamically favored due to an increase of the heat from optical power absorption resulting from the shift of $\omega_p$ in each resonator produced by charge transfer from the high frequency resonator to the low frequency resonator. Therefore a plasmoelectric device can be understood as a type of heat engine. The quantity of optically generated heat depends on the absorbed power from the incident radiation field, so that the magnitude of the plasmoelectric potential favoring charge transfer is dependent on the electron density-dependent absorption cross section, $C_{abs}(n)$, of each electrically coupled nanostructure.

To introduce the explicit dependence on electron density, $n$, in the absorption cross section, $C_{abs}$, we assume the complex dielectric function of each plasmonic resonator depends on the plasma frequency, $\omega_p$, according to a simple Drude model, with $\omega_p \propto n^{1/2}$ [12]. This strategy is consistent with other work that examined carrier density-dependent plasmon shifts, for example in doped semiconductors, electrochemical cells, or at metal surfaces during ultrafast pump-probe measurements.[13-16] A more detailed derivation is provided in the supporting information (SI). Figure 2 shows the change of $C_{abs}$ as $n$ is varied for a 10 nm radius Ag sphere surrounded by a dielectric matrix with refractive index n=2. Increasing or decreasing the electron density can increase the absorption of the particle at shorter or longer incident wavelengths, respectively.

The magnitude of the plasmoelectric potential depends on $C_{abs}(n)$ for a given radiation environment and thus can be deduced from a simple thermodynamic argument. We consider a system perturbed from equilibrium by an incident optical power density, $I_\lambda$. In steady state, this radiation gives rise to a Helmholtz free energy per unit volume available to the system, $F_{in}$, that can do work on the plasmoelectric device. The system can lower Helmholtz free energy by absorbing free energy from the radiation field, raising the temperature and entropic heat, $TS_R$, of the resonators. A spectral shift of the absorption maxima of the plasmoelectric device into resonance with the incident optical frequency provides a significant decrease of the free energy of the radiation field by increasing the quantity of heat absorbed by the resonator. However, a spectral shift requires a change of electron density in each of the coupled resonators and therefore increases the electrochemical free energy of the resonators, $F_R$. Charge transfer between resonators increases the internal energy of the resonators, $U_R$, by increasing both $TS_R$ and $F_R$ compared to an equilibrium configuration of uncharged resonators. In accordance with the 2nd law of thermodynamics, we solve for the device configuration that minimizes the total free energy of the system, $F_{in} - U_R + F_R$, with respect to the change in electron density, $n$, of each of the resonators. As derived in supplemental equations (S2)-(S7), the minimum total Helmholtz free energy corresponds to a configuration of the resonators where

$$\frac{dF_R}{dn} = \frac{d\overline{TS}_R}{dn} \qquad (1)$$

Here, $\overline{TS}_R$ is the equivalent internal entropy and temperature that an electrically neutral resonator with the same $C_{abs}(n)$ would exhibit under steady-state irradiation if no energy went to electrochemical work. An expression for $\overline{TS}_R$, based on an analysis of the Stefan-Boltzmann law, is provided in the supplemental information (equations S8-S13). A plasmoelectric device can increase system entropy with any transfers of charge density that increase the absorbed light energy, $\overline{TS}_R$, by more than the work, $F_R$, required to generate such an absorption cross section.

Equation (1) is a central result for understanding plasmoelectric behavior. This expression describes the thermodynamic favorability of charge transfer between resonators. Thus in general, the minimum free energy of the system corresponds to a configuration of charged, non-neutral particles. Equation (1) also defines the electrochemical potential of each resonator at the thermodynamically favored value of $n$. The condition is summarized for the calculated optical response for spherical Ag nanoparticles in Fig. 3 in the limit where no power is sent to a circuit load. The plasmoelectric effect may be interpreted as a shift in the internal electrochemical potential or Fermi level, $\epsilon_F$, of a plasmonic nanostructure under incident radiation that is analogous to "doping" or "gating" of a semiconductor. A decrease or increase of the electron density in a single resonator can minimize the total free energy of the system based on the relationship between the incident frequencies and the plasmon resonances of the nanoparticles.

Examination of the $C_{abs}(n)$ in Fig. 2 suggests several different strategies for power conversion via the plasmoelectric effect. For each scenario, we impose the constraint that the resonators are isothermally coupled, so that there is no additional thermoelectric potential across the device due to a thermal gradient between the resonators [17]. First, consider two resonators with different structures, e.g. particle radii, or surrounded by a different dielectric matrix so they exhibit absorption maxima at distinctly different frequencies. The resonators are chosen so that the upper left (red) quadrant of a plot of $C_{abs}(n)$ like that in Fig. 2, describing the low frequency resonator, overlaps in the same wavelength range as the lower right quadrant of the similar plot describing the high frequency resonator.

Then, incident radiation with wavelengths contained in the overlap region promotes charge transfer from the high frequency resonator to the low frequency resonator, and the absorption maxima of both resonators shift towards a frequency in the overlap region because of this charge transfer, as depicted in Fig. 1a. A thermodynamic analysis of the optical responses of individual resonators in this geometry is displayed in column (a) and (b) of Fig. 3. Irradiation at wavelengths predominantly outside the maximum absorption wavelength for the neutral particles promotes charge transfer in the opposite direction, for an identical device geometry. If charge transfers from the low frequency resonator to the high frequency resonator, the absorption maxima of both resonators shift farther apart.

Alternatively, two resonators with identical plasmon resonant absorption maxima but illuminated at different optical wavelengths can be electrically coupled. The first resonator only receives radiation at wavelengths shorter than the plasmon resonance wavelength for neutral particles, while the second resonator only receives radiation at wavelengths longer than the plasmon resonance wavelength for neutral particles. In this circumstance, a plasmoelectric current will flow from the second resonator to the first. A thermodynamic analysis of this device arrangement is presented in column (a) and (c) of Fig. 3. This approach, with illumination at two different incident wavelengths on each of two electrically coupled 10 nm radii Ag particles, is the scenario we consider with current-voltage modeling. We analyze this device geometry because it best describes the characteristic power conversion efficiency for a single type of plasmonic nanoparticle resonator structure.

We now solve for the current-voltage response. Figure 4 shows a device schematic and equivalent circuit of the modeled geometry, corresponding to a 10 nm Ag sphere irradiated at 550 nm (resonator A) electrically coupled via a non-resonant conductor to a 10 nm Ag sphere irradiated at 500 nm (resonator C) both under a radiation intensity of 1 kW m$^{-2}$ and surrounded with a dielectric matrix with refractive index n=2. We first express equation (1) in terms of an electrochemical potential defined by the change of the Fermi level, $\Delta\epsilon_F$, of an electron gas

$$\frac{dF_R}{dn} = \Delta\epsilon_F = \frac{d\overline{TS}_R}{dn} \qquad (2)$$

where $\Delta\epsilon_F = \epsilon_F(n) - \epsilon_F(n_{Ag})$ with $\epsilon_F$ given by the Fermi function. The value of $n$ that satisfies equation (2) defines the steady-state electrochemical potential on each resonator. These differential relationships are summarized in Fig. 3 in the limit where no power is sent to a circuit load. The total electrostatic potential, $\phi_{tot}$, induced by the plasmoelectric effect across both resonators equals the difference between the potentials on each individual resonator, if the resonators are isothermally coupled.

$$\phi_{tot} = (\Delta\epsilon_F)_A - (\Delta\epsilon_F)_C \qquad (3)$$

We note that any steady-state temperature gradient between resonators would produce an additional contribution to the thermoelectric potential. Based on column (a) and (c) of Fig. 3, we expect that $\phi_{tot} \approx 500$ mV for this device geometry at open circuit, i.e. when no power is sent to a circuit load, neglecting any temperature gradient. The magnitude of $\phi_{tot}$ decreases when power is sent to a circuit load because of the consequent decrease of internal temperature, as expressed in supplemental equations (S10) and (S11).

According to Fig. 4b, if the internal resistance of the resonator circuit, $R_i$, is sufficiently small, then the charge transfer induced by the plasmoelectric effect produces a voltage across the load, $V_{load}$, equivalent to the potential across the resonators, $\phi_{tot}$. The power lost to the circuit load, $P_{load}$, relates to this voltage drop and the magnitude of the current through the load, $\mathcal{I}_{load}$, according to Ohm's law.

$$P_{load} = V_{load} \cdot \mathcal{I}_{load} \qquad (4)$$

We see that the magnitude of the plasmoelectric potential depends on the power dissipated in the load through the temperature dependence of equation (2), but this potential also defines the power dissipated in the load through equation (4). To solve these coupled equations for the current response, we require an independent expression relating $P_{load}$ with the device response via the $C_{abs}(n)$.

Conservation of energy provides the necessary constraint on $P_{load}$ to solve these equations. Under steady-state conditions, conservation of energy requires that all of the power absorbed by the plasmonic resonators is either sent to a load or re-emitted. This absorbed power depends on the $C_{abs}(n)$ of the device.

$$P_{abs} = I_\lambda \cdot C_{abs}(n) = P_{emit} + P_{load} \qquad (5)$$

For the device configuration when no absorbed power is used to transfer charge between resonators, both resonators exhibit a neutral charge configuration, and no potential develops across the load. Equation (5) reduces to

$$P_{abs}(n_{Ag}) = I_\lambda \cdot C_{abs}(n_{Ag}) = P_{emit}(n_{Ag}) \qquad (6)$$

We subtract this equality from both sides of equation (5) to give

$$P_{abs}(n) - P_{abs}(n_{Ag}) = P_{emit}(n) - P_{emit}(n_{Ag}) + P_{load}(n) \qquad (7)$$

This form emphasizes that only the increased fraction of the absorbed power that results from charge transfer can be used to maintain the non-equilibrium configuration of internal electrochemical free energy. Then, the maximum power that can be sent to a load is the difference of absorbed power at the thermodynamically favored configuration compared with a neutral configuration. For the specific incident radiation profile, we solve for the thermodynamically favored resonator potential, $\phi_{tot}$, and the resulting current through the load, $\mathcal{I}_{load}$, when $P_{load}$ takes the limiting minimum value of zero, corresponding to open circuit, through the maximum possible value, $P_{load} = P_{abs}(n) - P_{abs}(n_{Ag})$, corresponding to short circuit.

The resulting current and voltage values for this range of $P_{load}$ are displayed as a power curve in Fig. 4c. The calculated value of the open circuit voltage, $V_{OC}$ = 454 mV, is approximately 50 mV lower than the value estimated above of $\phi_{tot} \approx 500$ mV. This difference reflects the constraint that both resonators are isothermally coupled as modeled here, and that as a pair, they adopt a steady-state temperature that is the average temperature of the respective individual resonators displayed in column (a) and (c) of Fig. 3. We determine the optical power conversion efficiency by dividing the value of $\mathcal{I}_{load} \cdot V_{load}$ at each value of $V_{load}$ by the absorbed power from the radiation field when the $C_{abs}(n)$ adopts the value of $n$ corresponding to that potential. The highest conversion efficiency for the device may not correspond to operation at the maximum power point of the current-voltage response, because the absorbed power depends strongly on the potential adopted by the resonators. For this geometry the highest conversion efficiency of 14.3 %, corresponds to a voltage of $V_{load}$ = 241 mV, which is 46 mV lower than the voltage that maximizes power through the load, at $V_{load}$ = 287 mV.

Finally, we comment on strategies for broadband optical power conversion using the plasmoelectric effect. To extend our analysis to broadband incident radiation, it is necessary to consider the coupled equations (2) and (5) when the internal heat of the resonators depends on a specific incident spectral radiation profile. Because the maximum power that can be sent to a load depends on the difference $P_{abs}(n) - P_{abs}(n_{Ag})$, an efficient device will maximize the increase of absorbed power for some change of electron density. A proposed broadband power conversion scheme is presented in Fig. 5. We imagine a device composed of identical arrays of coupled resonators, with the absorption maxima for several neutral particle plasmonic resonators spanning the incident radiation spectrum. A passive optical splitting element sends the incident broadband radiation to each resonator array with a radiation profile optimized to maximize the plasmoelectric potential across the entire device. In the limit that the optical splitting element and resonator line widths restrict absorption of any of the incident radiation to configurations that can maximize absorption via charge transfer, then all of the absorbed radiation power could in principle be sent through a circuit load. However, because a plasmoelectric device is fundamentally a type of heat engine, we expect that the thermodynamic power conversion efficiency for such an idealized arrangement can approach the Carnot limit. The relevant temperature difference for the Carnot analysis corresponds to the steady-state temperature of the resonator array at the thermodynamically favored value of electron density versus the corresponding temperature for a neutral particle in a dark ambient.

In this Letter, we have proposed the plasmoelectric effect as an approach for conversion of optical power into DC electrical power via resonant absorption in plasmonic nanostructures, and we have outlined several experimentally testable predictions that distinguish this phenomenon from the more familiar thermoelectric and photovoltaic effects [17, 18]. Although we considered the specific case of Ag nanoparticles under irradiation at discrete wavelengths, our approach can be extended to a variety of material systems, resonator geometries, and radiation environments. Besides employing the plasmoelectric effect for optical power conversion or wireless power transfer, we believe the phenomenon may allow a new class of optoelectronic devices. For example, the plasmoelectric effect could be employed to modulate the absorption cross section at specific frequencies for optical switching applications. With an appropriate circuit design, a broadband device like that in Fig. 5 could produce an electrical signal that is characteristic of the incident spectrum, enabling subwavelength spectrometers with electronic readout. Fundamentally, the plasmoelectric mechanism provides active control of the Fermi level of a metal using incident radiation. This may enable development of new types of all-metal optoelectronic devices by replacing the usual function of doped or gated semiconductor components with metal nanostructures that are optically excited off-resonance, an approach that is facilitated by the remarkable spectral tailorabilty of plasmonic nanostructures.

Acknowledgements:

The authors gratefully acknowledge support from the Department of Energy, Office of Science under contract number DE-FG02-07ER46405. Helpful discussions with E. Kosten, V. Brar, D. Callahan, M. Deceglie, A. Leenheer, and J. Fakonas are gratefully acknowledged.

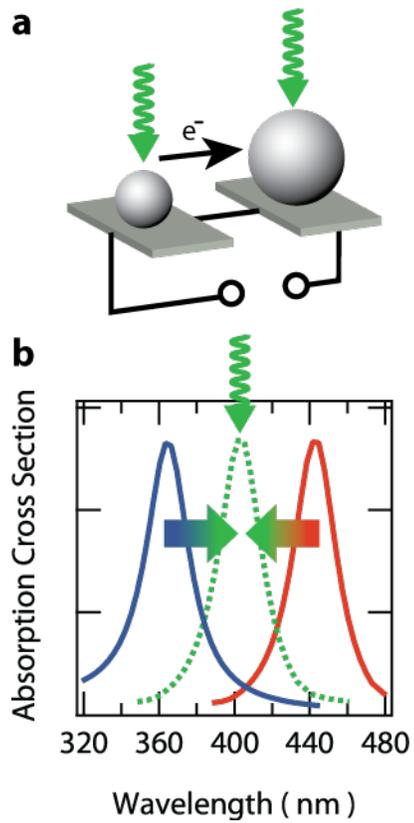

**Figure 1** Schematic of the plasmoelectric effect. (a) Electrically coupled plasmonic resonators (b) with resonant absorption maxima at distinct frequencies (blue or red) are irradiated at an intermediate frequency (green). Charge transfer between resonators is thermodynamically favored because of the increase of heat due to absorption that accompanies the consequent spectral shifts.

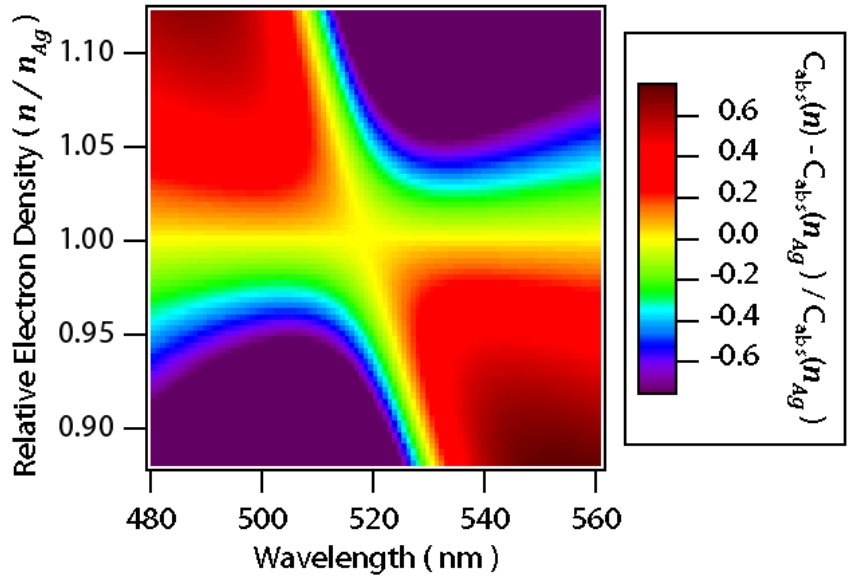

**Figure 2** Electron density-dependent absorption cross section. The relative change of the absorption cross section, $C_{abs}$, as a function of electron density, $n$, is compared with the absorption cross section corresponding to the electron density, $n_{Ag}$, of a neutral silver resonator. The modeled geometry is a 10 nm radius sphere surround by a medium with refractive index n=2.

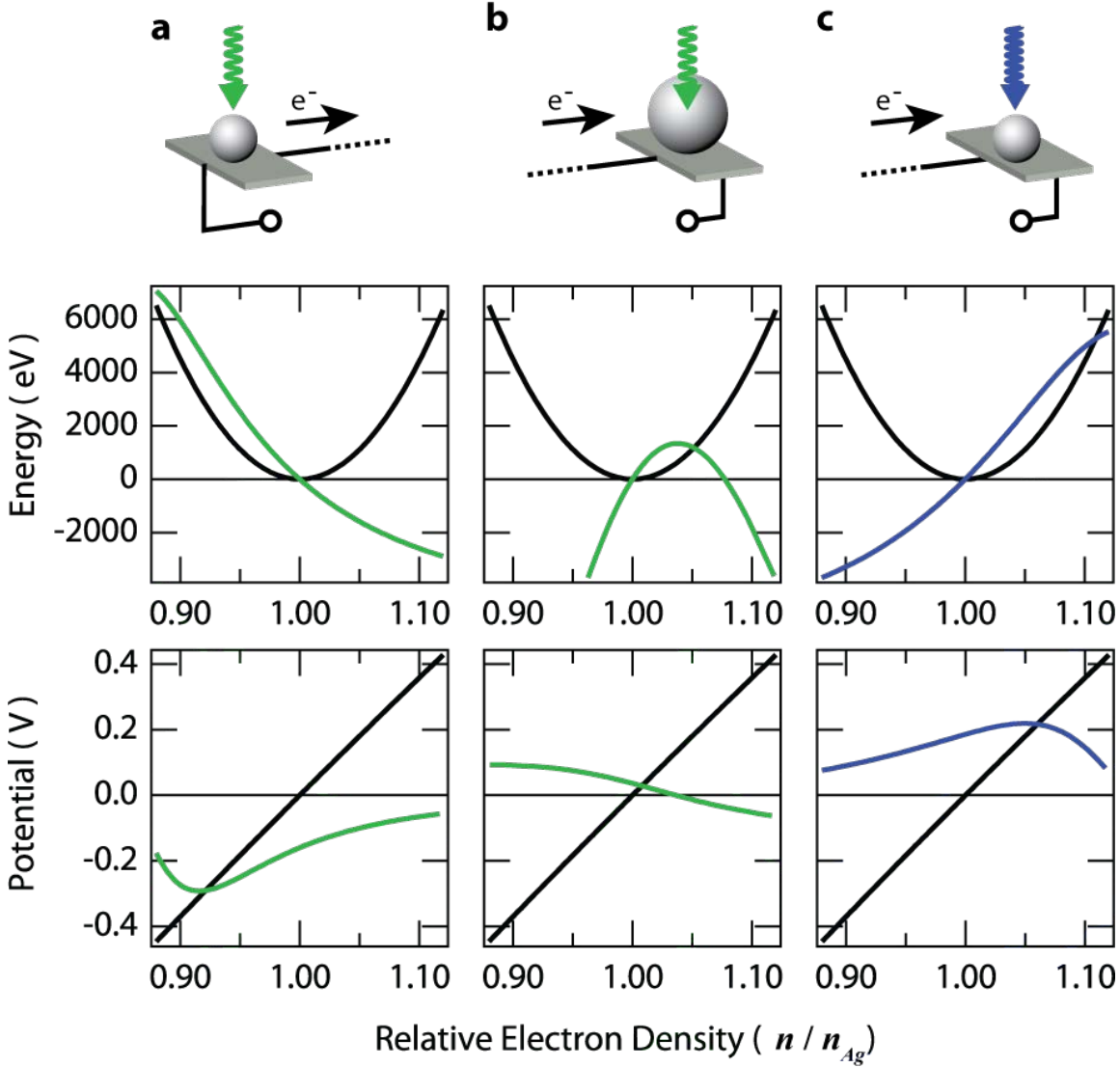

**Figure 3** Thermodynamic analysis of plasmoelectric particle configurations. When no power is sent to a circuit load, the radiation profile and resonator geometry define the optical response of (a) a 10 nm radius Ag sphere under 1 kW m$^{-2}$ excitation at 550 nm (green), (b) a 22 nm radius Ag sphere under the same irradiation, or (c) a 10 nm radius Ag sphere under 1 kW m$^{-2}$ excitation at 500 nm (blue). **Middle:** The electrochemical free energy, $F_R$ (black), associated with charge transfer is plotted with the steady-state internal heat from absorption, $\overline{TS}_R$ (green or blue), for each resonator system as a function of electron density. **Bottom:** The differential heat from absorption, $\frac{d\overline{TS}_R}{dn}$ (green or blue), is plotted with the differential free energy, $\frac{dF_R}{dn}$ (black), where $\frac{dF_R}{dn} = \Delta\epsilon_F$. The minimum free energy of the system, when $\frac{dF_R}{dn} = \frac{d\overline{TS}_R}{dn}$, defines the electrochemical potential of the resonator induced by the plasmoelectric effect and the thermodynamically favored $C_{abs}(n)$. An efficient device electrically couples resonators that maximize the total potential difference, such as resonators (a) and (b) or resonators (a) and (c).

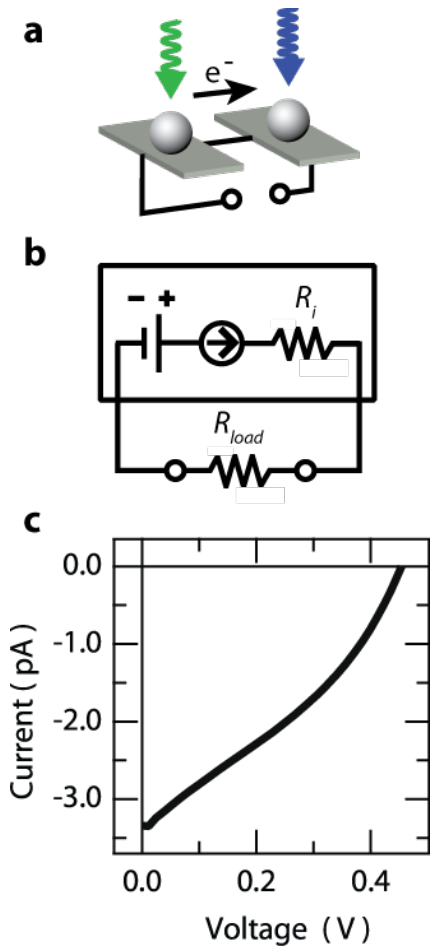

**Figure 4** Modeled current-voltage response. (a) Two 10 nm Ag spheres surrounded by n=2 dielectric are optically isolated, and irradiated at 1 kW m$^{-2}$ with 550 nm (green) or 500 nm (blue) radiation (b) The equivalent circuit of the solved geometry is an ideal voltage source in series with an ideal current source, an internal resistance, $R_i$, and a load, $R_{load}$. (c) The current-voltage plot exhibits a maximum optical power conversion efficiency of 14.3 % for device operation at 241 mV. The short circuit current is 343 pA, and the open circuit voltage is 454 mV.

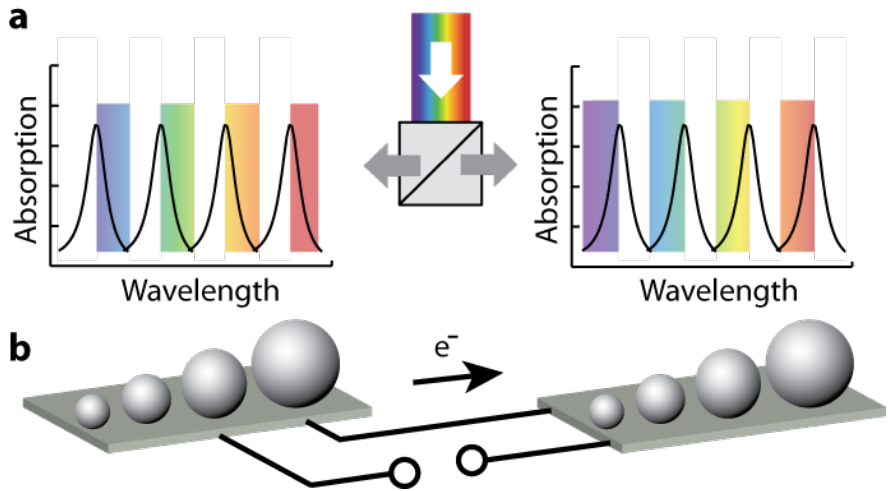

**Figure 5** Broadband power conversion scheme. (a) A passive optical splitting element separates incident broadband radiation to maximize the plasmoelectric potential across (b) arrays of plasmonic resonators with absorption maxima spanning the incident radiation spectrum.

**SUPPLEMENTAL:**

**Complex Dielectric Function**

To describe an electrically neutral Ag plasmonic resonator, we apply a 6th-order, multiple oscillator Lorenz-Drude model of the complex dielectric function of silver, fit to data from the Palik Handbook. Rakic and coworkers outline the method used here [1]. This dielectric function accurately reproduces the observed extinction spectra of spherical silver nanoparticles when input into the exact analytic solutions to Maxwell's equations provided by Mie theory [2]. To introduce the explicit dependence on changes of electron density, we assume that all terms in the dielectric function that depend on the bulk plasma frequency, $\omega_p$, depend on electron density, $n$, according to a simple Drude model,

$$\omega_p = \left( \frac{n \cdot e^2}{\varepsilon_o \cdot m_e^*} \right)^{1/2} \qquad (S1)$$

where $e$ is the electron charge, $\varepsilon_o$ is the permittivity of free space, and $m_e^*$ is the electron effective mass [3]. This strategy is consistent with other work that examined carrier density-dependent plasmon shifts, for example in doped semiconductors, electrochemical cells, or at metal surfaces during ultrafast pump-probe measurements [4-7].

**Thermodynamic Device Model**

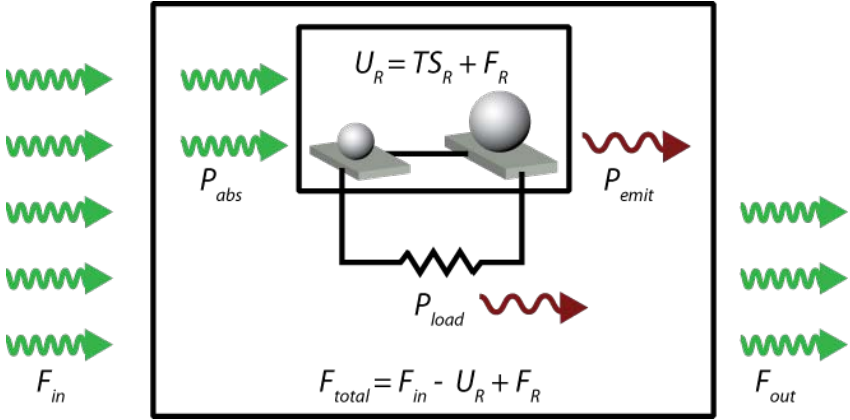

**Figure S1** Thermodynamic device model. The incident radiation field comprises a free energy available to the system, $F_{in}$. A spectral shift of the absorption maxima of the device into resonance with the radiation lowers the Helmholtz free energy by increasing the heat of both resonators, $TS_R$. However, a spectral shift also increases the Helmholtz free energy of the resonators, $F_R$, because of the change of electron density required to alter the resonance frequencies.

We consider a system perturbed from equilibrium by an incident optical power density, $I_\lambda$. In steady-state, this radiation gives rise to a Helmholtz free energy per unit volume available to the system, $F_{in}$, that can do work on the plasmoelectric device, as depicted in Fig. S1. The system can lower its overall Helmholtz free energy by absorbing free energy from the radiation field, raising the temperature and entropic heat, $TS_R$, of the resonators. A spectral shift of the absorption maxima of the plasmoelectric device into resonance with the incident optical frequency provides a significant decrease of the free energy of the radiation field. However, spectral shifts require a change of the electron density in each of the coupled resonators and therefore increase the electrochemical energy of the resonators, $F_R$. In accordance with the 2nd law of thermodynamics, we solve for the device configuration that minimizes the total free energy of the system with respect to changes of the electron density, $n$, of the resonators.

Under steady-state conditions, conservation of energy requires that all of the power absorbed by the plasmonic resonators is either sent to a load or re-emitted. This absorbed power depends on the $C_{abs}(n)$ of the device.

$$P_{abs} = I_\lambda \cdot C_{abs}(n) = P_{emit} + P_{load} \qquad (S2)$$

The specific $C_{abs}(n)$ the device adopts in steady state defines the total internal energy, heat, and free energy of the resonators. These relations are derived explicitly below, and reflect the internal temperature and charge density associated with a given configuration. For changes of electron density,

$$\frac{dU_R}{dn} = \frac{dTS_R}{dn} + \frac{dF_R}{dn} \qquad (S3)$$

If a resonator were always electrically neutral, an increase of internal energy due to increased absorption would be converted entirely into heat resulting in an increase of temperature. However, for this system the increase of heat is moderated by the necessary fraction that is converted into electrochemical free energy.

$$\frac{dTS_R}{dn} = \frac{d\overline{TS}_R}{dn} - \frac{dF_R}{dn} \qquad (S4)$$

Here, $\overline{TS}_R$ is the equivalent internal entropy and temperature that an electrically neutral resonator with the same EDDACS would exhibit under steady-state irradiation if no energy went to electrochemical work. An expression for $\overline{TS}_R$, is provided below. A plasmoelectric device can increase entropy with any charge configurations that increase the absorbed light energy by more than the work required to generate such an absorption cross section.

It is clear that the total free energy in the system depends on the fraction of incident radiation that is converted into the internal energy of the resonators and the fraction of that internal energy that is further converted into the electrochemical free energy associated with charge transfer.

$$F_{total} = F_{in} - U_R + F_R \qquad (S5)$$

Because the incident power has no dependence on the charge configuration of the resonators, the following expression describes the minimum free energy of the system with respect to changes of electron density:

$$\frac{dF_{total}}{dn} = 0 = \frac{dF_R}{dn} - \frac{dU_R}{dn} \qquad (S6)$$

Substitution of equations (S3) and (S4) shows this condition corresponds to a configuration of the resonators where

$$\frac{dF_R}{dn} = \frac{d\overline{TS}_R}{dn} \qquad (S7)$$

**Steady State Internal Heat, $\overline{TS}_R$**

Here, we determine the steady-state internal temperature and entropy of a nanostructure while it is irradiated at a single wavelength and intensity, $I_\lambda$, for a specific $C_{abs}(n)$. This analysis is extended to larger incident bandwidths by accounting for the spectral power absorbed for some finite wavelength range. We consider a device like that depicted in Fig. S1. A two-temperature model describes the power flow between the conduction electrons, phonons, the circuit load, and the ambient environment during irradiation [2]. In general, the rate of energy exchange between electrons and phonons depends on the electron-phonon coupling constant, $g$, the respective lattice and electronic heat capacities, $C_l$ and $C_e$, and the relative temperature difference [8]. We make the following assumptions regarding the device behavior, summarized in the coupled equations (S8) and (S9).

$$C_l \frac{dT_l}{dt} = g(T_e - T_l) - P_{emmission} + P_{ambient} \qquad (S8)$$

$$C_e \frac{dT_e}{dt} = -g(T_e - T_l) + I_\lambda \cdot C_{abs}(n) - P_{load} \qquad (S9)$$

All of the power absorbed by the electrons, $I_\lambda \cdot C_{abs}(n)$, is emitted into the phonons, unless that power is lost to the circuit load, $P_{load}$. Therefore we do not consider resonant optical emission from excited electron-hole pairs, which occurs with an efficiency of less than $10^{-6}$ for metal nanoparticles [9]. We also assume that the phonons only exchange energy with the ambient environment directly via black body emission or absorption, in accordance with the Stefan-Boltzmann law. Under steady-state conditions, this gives the following expression for the electron and phonon temperature of a resonator,

$$T_e = \frac{I_\lambda \cdot C_{abs}(n) - P_{load}}{g} + T_l \qquad (S10)$$

$$T_l = \left(\frac{I_\lambda \cdot C_{abs}(n) - P_{load} + \sigma \varepsilon A_s T^4_{ambient}}{\sigma \epsilon A_s}\right)^{1/4} \quad (S11)$$

where $A_s$ is the surface area of the particle, $\varepsilon$ is the emissivity ($\varepsilon = 1$ for a perfect black body) and $\sigma = 5.67 \times 10^{-8}$ J s$^{-1}$ m$^{-2}$ K$^{-4}$ is the Stefan-Boltzmann constant.

The total steady-state internal heat of the resonator induced by absorption depends on the temperature-dependent thermal energy of both phonons and electrons.

$$\overline{TS}_R = T_l S_l + T_e S_e \quad (S12)$$

The high temperature lattice heat capacity of silver defines the increase of phonon energy with temperature. Our calculations use the reported value of $C_l = 2.44 \times 10^6$ J m$^{-3}$ K$^{-1}$ [10]. The lattice heat capacity does not depend explicitly on electron density, so the phonon energy of the resonator depends on electron density only through the dependence of $T_l$ on the $C_{abs}(n)$ expressed in equation (S11). However, the internal electronic heat is a function of the electron density, $n$, and electronic temperature, $T_e$, as described by the Sommerfeld model for the electronic heat capacity of metals [11].

$$T_e S_e = \frac{m_e k_B^2 V \pi^{2/3}}{\hbar^2 3^{2/3}} T_e^2 n^{1/3} \quad (S13)$$

Here, $V$ is the particle volume, $m_e$ is the electron rest mass, $\hbar$ is the reduced Plank constant, and $k_B$ is the Boltzmann constant. We see that increases of electronic temperature efficiently compensate for entropy losses associated with decreased electron density in a plasmoelectric device.